\input{aipcheck}

\documentclass[
    ,final            
  ]
  {aipproc}

\layoutstyle{8x11double}

\begin{document}

\title{Nucleosynthesis in Metal-Free and Metal-Poor Stars}



\author{Yong-Zhong Qian}{
  address={School of Physics and Astronomy, University of Minnesota, 
  Minneapolis, MN 55455}
}

\begin{abstract}
There have been a number of important recent developments in theoretical 
and observational studies of nucleosynthesis, especially regarding nucleosynthetic 
sources at low metallicities. Those selected for discussion here include the
origin of $^6$Li, the primary production of N, the $s$-process, and
the supernova sources for three groups of metals: (1) C to Zn
with mass numbers $A<70$, (2) Sr to Ag with $A\sim 90$--110,
and (3) $r$-process nuclei with $A\sim 130$ and above.
\end{abstract}

\maketitle

\section{Introduction}

Metals provide cooling in a star-forming gas, and the metallicity of the gas
may be related to the initial mass
function (IMF) of the stellar populations to be formed. In particular, it is considered that
only massive stars can be formed from primordial gas with zero metallicity.
Thus, metallicity can affect nucleosynthesis through its effects on the formation of 
different stellar populations as nucleosynthetic sources.
The initial metallicity of a star also has important effects on its structure and evolution,
and in the case of a massive star, on its explosion. These effects can influence
nucleosynthesis during stellar evolution and explosions. Finally, the initial metallicity 
of a star can affect nucleosynthesis directly.

The above effects of metallicity on nucleosynthesis are illustrated nicely by the case 
of very massive stars (VMSs) of $\approx 130$--$300\,M_\odot$ that give rise to
pair-instability supernovae (PI-SNe). First of all, it is
expected from recent simulations that due to the inefficient cooling by H$_2$ molecules,
the collapse of metal-free gas clouds results in the formation of stars of 
$\sim 10$--$10^3\,M_\odot$ (see e.g., \cite{abel,bromm} for reviews of earlier works
and \cite{yoshida,oshea,gao} for more recent studies). 
Without rotation, these massive metal-free stars 
experience very little mass loss and end their lives with essentially their initial masses
(e.g., \cite{baraffe}). 
A metal-free VMS of $\approx 130$--$300\,M_\odot$ encounters the pair instability 
following core He burning. It implodes and then explodes as a PI-SN that disrupts the 
entire VMS \cite{heger}. 
These PI-SNe have a distinct nucleosynthetic signature: the gross 
underproduction of the elements of odd atomic numbers such as Na, Al, P, Cl, K, Sc, 
V, Mn, Co, and Cu \cite{heger,umeda}. This is because the production of these elements 
requires a significant neutron excess but a metal-free VMS has no initial neutron excess 
and evolves too quickly to allow weak interaction processes such as $\beta^+$ decay to 
produce a significant neutron excess prior to its explosion.

The case of PI-SNe also nicely illustrates the interaction between theory and 
observations. The extensive data (e.g., \cite{cayrel1}) on extremely metal-poor stars with 
$-4<{\rm [Fe/H]}=\log({\rm Fe/H})-\log({\rm Fe/H})_\odot<-3$ show that
the abundances of the elements of odd atomic numbers relative to their neighboring 
elements in these stars are never as low as predicted for the products from 
PI-SNe. This suggests the following possibilities: (1) the VMS progenitors for PI-SNe
cannot be formed from metal-free gas; (2) although such progenitors were formed
and gave rise to PI-SNe, their nucleosynthetic contributions were overwhelmed by 
other sources that have less
severe underproduction of the elements of odd atomic numbers; or (3) although
metal-free VMSs of $\approx 130$--$300\,M_\odot$ were formed, their initial
masses were reduced during their evolution due to e.g., rotation \cite{meynet1}. 
Another interesting development is the recent report of possible detection of a modern
PI-SN \cite{nsmith}. Clearly, further theoretical and observational studies are required
in order to resolve whether VMS progenitors for PI-SNe can be formed at zero
or any metallicity and how such stars evolve and make metals when more 
physical ingredients such as rotation are included in the model.

\section{Effects of Rotation on Nucleosynthesis in Metal-Free and 
Metal-Poor Stars}

While not easy to treat, rotation is a common property of stars and can have
crucial effects on nucleosynthesis. To illustrate these effects, we discuss two 
examples: nucleosynthesis associated with jet-powered explosions of
metal-free stars of 300--$1000\,M_\odot$ and primary production of N by 
massive as well as low and intermediate mass stars with rotation at low 
metallicities.

In the absence of rotation, metal-free stars of 300--$1000\,M_\odot$ collapse into 
black holes at the end of their lives without ejecting any material. With significant
rotation, an accretion disk may form around the black hole produced by the core
collapse of such a star and the jet associated with the disk may power an explosion. 
The nucleosynthetic products ejected by jet-powered explosions of metal-free
stars of 500 and $1000\,M_\odot$, respectively, were studied in \cite{ohkubo}
assuming a range of opening angles, velocities, energies, and amounts of
ejecta for the jets. It was found that the yield patterns for particular sets of jet 
parameters (especially models B-1 and B-2 in \cite{ohkubo}) are in agreement 
with the abundance ratios [O/Fe], [Ne/Fe], [Mg/Fe], [Si/Fe], and [S/Fe] observed
in the intracluster medium and in the gas of the central region of M82. 
The same yield patterns
are also in approximate agreement with the abundance ratios [Mg/Fe], [Si/Fe],
[Ca/Fe], [Cr/Fe], [Co/Fe], [Ni/Fe], and [Zn/Fe] observed in extremely metal-poor 
stars with $-4<{\rm [Fe/H]}<-3$ although the models give 
significantly or much lower values for
[O/Fe], [Ti/Fe], and [Mn/Fe]. In addition, the abundance ratio ${\rm [C/Si]}\sim -0.77$ 
inferred from observations for the intergalactic medium \cite{aguirre}, 
which cannot be accounted 
for by regular SNe and was attributed to a mixture of products from PI-SNe and
regular SNe (e.g., \cite{qw1}), may be provided by jet-powered explosions of metal-free
stars of 300--$1000\,M_\odot$ \cite{ohkubo}. 
The intermediate-mass black holes left behind
by such stars may serve as seeds for the supermassive black holes observed
in the galactic nuclei of the present universe. The association of the nucleosynthetic 
contributions from metal-free stars of 300--$1000\,M_\odot$ with the net inventory of 
intermediate-mass black holes remains to be explored (see \cite{qw2} for discusssion
of a similar association based on attributing ${\rm [C/Si]}\sim -0.77$ in the intergalactic 
medium to a mixture of products from PI-SNe and regular SNe).

Observations \cite{mspite} show that the abundance ratio N/O is rather high with 
a large scatter but no clear trend at low metallicities, 
for which massive stars were the predominant sources for metals.
As O is a primary element produced by massive stars, this requires a primary 
mechanism for producing N also by these stars. For metal-poor
stars of low to high masses, primary production of N can
be achieved by mixing the C from He burning into the H burning shell, where N
is produced by the CN cycle. Adequate mixing may be induced if a star starts on
the main sequence with the amount of rotation comparable to that
observed in present-day stars.
The transport of the newly-synthesized C into the H burning shell is especially 
efficient for rotating metal-poor stars because their lower mass loss 
rates and greater compactness result in steeper angular velocity gradients, 
which in turn cause more mixing due to the shear instability \cite{meynet2}.
Calculations of evolution and nucleosynthesis in 
metal-poor stars of low to high masses with reasonable initial rotation show 
that stars of $\sim 5$--$7\,M_\odot$ have the largest absolute N yields and
the dominant sources for primary N when weighted by the Salpeter IMF
are stars of $\sim 2$--$5\,M_\odot$ \cite{meynet2,meynet3}. These results can
account for the general trend for the evolution of the N/O ratio with metallicity
\cite{meynet3,chiappini}, in particular the high N/O ratios at low
metallicities for which massive stars were the predominant sources for N 
through primary production enhanced by rotationally-induced mixing.
Further, rotation can induce severe mass loss from extremely metal-poor 
stars of high masses due to the enhanced surface metallicity following 
the dredge-up of newly-synthesized metals \cite{meynet1}. 
This suggests that extremely metal-poor VMSs may suffer so much mass
loss that they could not produce PI-SNe. Finally, in addition to N, 
the abundance ratios of C, O, Na, Mg,
and Al relative to Fe achieved in the envelope of rotating metal-poor stars of
2--$7\,M_\odot$ are in accordance with the ratios observed in non-evolved
carbon-enhanced metal-poor stars \cite{meynet1}.

\section{Nucleosynthesis at Low Metallicities}

Now we turn to a general discussion of nucleosynthesis at low metallicities,
which covers the sources for the fragile $^6$Li, the intermediate-mass elements
from C to Zn, and the heavy elements usually attributed to the $s$-process and 
the $r$-process.

\subsection{Sources for $^6$Li}

The Spite plateau \cite{spite} of 
$\log\epsilon({\rm Li})=\log({\rm Li/H})+12=2.05\pm0.15$
at $[{\rm Fe/H}]<-1.5$ is well established from observations of metal-poor
stars in the Galactic halo by different groups. It is widely accepted that
this Li represents the $^7$Li produced in the big bang. However, for
the baryon-to-photon ratio measured by the recent cosmic microwave
background experiment WMAP (e.g., \cite{spergel}), the theory of
standard big bang nucleosynthesis (SBBN) predicts
$\log\epsilon(^7{\rm Li})=2.65\pm0.10$ (e.g., \cite{cyburt}), which is
higher than the Spite plateau by $\sim 0.5$ dex. As $^7$Li is destroyed 
at relatively low temperature,  a plausible explanation of this discrepancy 
is that the surface abundances of this isotope in metal-poor stars
have been depleted by processes in the surface region.
A scenario where this depletion is achieved approximately independent 
of e.g., the surface temperature of stars was discussed in \cite{richard}.

Accepting that the Spite plateau results from nearly uniform depletion
of the primordial $^7$Li, we are facing another problem with the origin
of $^6$Li detected in stars on the plateau.
The presence of $^6$Li introduces a slight additional asymmetry into 
the intrinsically asymmetric line profile of $^7$Li
and the spectral analysis to extract the $^6$Li/$^7$Li ratio is
rather difficult. Nevertheless, at least the detection of $^6$Li
in HD~84937 has been confirmed
by several groups (e.g., \cite{smith1,hobbs,cayrel}). The measured
isotopic ratio $^6$Li/$^7$Li~$=0.052\pm0.019$ \cite{cayrel}
is orders of magnitude larger than the value
$^6$Li/$^7$Li~$\sim 10^{-5}$ resulting from SBBN.
A more recent observational study of 24 metal-poor halo dwarfs has
detected $^6$Li in 9 stars at the $\geq 2\sigma$ significance level and
suggests a $^6$Li plateau of $\log\epsilon(^6{\rm Li})\approx 0.8$
corresponding to $^6$Li/$^7$Li~$\sim 0.05$ \cite{asplund}. 
As $^6$Li is much easier to destroy than $^7$Li,
this implies that there must be sources at low metallicities
that can produce $^6$Li much more efficiently than SBBN.

One possible mechanism for producing $^6$Li at low metallicities
relies on cosmic rays. Some of the primordial $^4$He nuclei 
($\alpha$-particles) can be accelerated into energetic 
cosmic rays by shocks from e.g., structure formation and SNe.
The fusion of the $^4$He nuclei in the cosmic rays and those in
the general medium produces $^6$Li through e.g., the reaction
$\alpha+\alpha\to{^6{\rm Li}}+d$. By considering a cosmological
component of cosmic rays possibly accelerated by the first stars,
it was shown that the $^6$Li plateau could be produced 
(e.g., \cite{rollinde}). In general, cosmic-ray production of $^6$Li 
can proceed through both fusion of $\alpha$-particles and spallation
of C, N, and O nuclei by protons and $\alpha$-particles. It is 
expected that the fusion channel dominates at low metallicities.  
At metallicities close to the solar value, the $^6$Li abundances 
are dominated by contributions from cosmic-ray spallation, which
is the only channel to produce Be. The ratio $^6$Li/Be~$\approx 6$
from meteorites is characteristic of the relative production of
these two nuclei by spallation. This ratio is much lower than the value 
$^6$Li/Be~$\approx 40$ observed in HD~84937 \cite{smith2}, which
is consistent with the dominant additional contributions to $^6$Li from 
fusion of $\alpha$-particles at low metallicities.

The main reaction for producing $^6$Li in SBBN is
$\alpha+d\to{^6{\rm Li}}+\gamma$, which is a very inefficient
electromagnetic quadrupole reaction (the dipole of the reacting
nuclei vanishes as they have the same charge-to-mass ratio).
On the other hand, the production of $^6$Li can be greatly enhanced 
in non-SBBN. For example, the hadronic 
decay of supersymmetric particles produce energetic nucleons,
which in turn can produce $^3{\rm H}$ and $^3{\rm He}$ through 
spallation of $^4{\rm He}$. If such decay occurs $\sim 10^3$~s after 
the big bang, $^6$Li can be produced through reactions such as 
$\alpha+{^3{\rm H}}\to{^6{\rm Li}}+n$
and $\alpha+{^3{\rm He}}\to{^6{\rm Li}}+p$. 
In fact, for the appropriate particle properties, it is possible
to enhance the production of $^6$Li and destroy $^7$Li 
simultaneously without affecting the yield of D very much 
(e.g., \cite{jedamzik}). In this case, the Spite plateau of
$^7$Li and the accompanying plateau of $^6$Li observed in 
metal-poor stars would be interpreted simply as the true primordial 
abundances produced by non-SBBN without alteration
by stellar processing or additional contributions from other sources.
The above non-SBBN scenario is attractive in that it solves 
two problems of the Li isotopes at the same time. However, the
required physics input, while plausible, is not yet established.

\subsection{The $s$-process}

The main $s$-process producing $^{88}$Sr and heavier nuclei 
occurs in asymptotic giant branch (AGB) stars. In the current model,
a $^{13}$C pocket is formed via the reaction sequence 
${^{12}{\rm C}}+p\to{^{13}{\rm N}}+\gamma$ followed by
${^{13}{\rm N}}\to{^{13}{\rm C}}+e^++\nu_e$ after a small quantity
of protons are mixed into the region between the He and H burning shells.
Neutrons produced by the reaction
${^{13}{\rm C}}+\alpha\to{^{16}{\rm O}}+n$ are then captured by
Fe seed nuclei to make $s$-process nuclei ($s$-nuclei). As the efficiency
of the neutron source is considered to be independent of the
initial metallicity of the AGB star but the abundance of Fe seed nuclei 
directly depends on this metallicity, large neutron-to-seed ratios are
achieved in metal-poor AGB stars \cite{gallino}. Therefore, major
production of Pb is expected from the $s$-process at low metallicities. 

While the $s$-process may occur in AGB stars of higher masses,
the dominant sources for $s$-nuclei are low-mass
AGB stars of $\sim 1$--$3\,M_\odot$ when the Salpeter IMF is
taken into account. As low-mass stars have long evolution timescales,
the $s$-process contributions to the general interstellar medium (ISM) 
are negligible at low metallicities, say [Fe/H]~$<-1.5$. However,
in a binary system consisting of two metal-poor stars of low but different 
masses, the primary star of higher mass would go through the AGB phase 
in due time and contaminate the surface of the unevolved secondary star 
with AGB products through mass transfer. The secondary star would then 
be observed to have extremely high enrichments of 
$s$-nuclei and other AGB products, especially Pb and C, but very low
abundances of non-AGB products such as Fe. A number of these
so-called Pb stars have been observed and their abundance patterns
of Zr, La, Ce, Pr, Nd, Sm and Pb can be fitted by calculations based
on the above $s$-process model (e.g., \cite{vaneck}).

There are also a number of metal-poor stars observed to have extremely 
high enrichments of C and Pb but their overall abundance patterns
of Ba and above appear to be a mixture of 
$s$-process and $r$-process products
(e.g., \cite{hill,cohen,jonsell}. It was proposed that these so-called $s+r$
stars were initially the secondary members of binary systems. The primary 
member of such a system first went through the AGB phase, transferring 
$s$-nuclei and other AGB products onto the surface of the 
secondary member. The white dwarf left behind by the AGB evolution
then accreted back some of the material and collapsed into a neutron star. 
The ejecta from this accretion-induced collapse (AIC) event contaminated 
the surface of the secondary member with the 
$r$-process nuclei ($r$-nuclei) produced 
in the event \cite{qw3}. This AIC scenario requires the primary member
to be rather massive so that the white dwarf produced by its AGB evolution
needs to accrete only a small amount of material to collapse.
It is also possible that the primary member is in a narrow mass range
($\sim 8\,M_\odot$) for which a core-collapse SN follows the AGB evolution
\cite{wanajo1}. In any case, both the $s$-process and the $r$-process in the 
above binary scenarios need much further investigation. For example,
the data on Ba and a number of rare-earth elements in CS~31062--052 
and LP~625--44 cannot be explained by a mixture of the solar $r$-process
abundance pattern and the $s$-process yield pattern calculated using 
the $^{13}$C pocket \cite{aoki}. The $r$-process is discussed in 
some detail below.

\subsection{Nucleosynthesis in massive stars}

The dominant sources for the intermediate-mass elements from C to Zn
at low metallicities are massive stars of $\sim 10$--$100\,M_\odot$,
which give rise to Fe core-collapse SNe at the end of their evolution. 
The elements from C to Al are mainly produced by hydrostatic burning 
during the pre-SN evolution, and the elements from Si to Zn are
mainly produced by explosive burning associated with the propagation
of the SN shock through the shells above the Fe core. While the production
of all these elements has some important dependence on the initial 
metallicity of the SN progenitor through e.g., mass loss, pre-SN density
structure, and details of the explosion, the yields of individual elements
are broadly similar
over a wide range of metallicities and have been calculated by several
groups (e.g., \cite{ww95,chieffi,tominaga}).

\subsubsection{Mass cut, fallback, and mixing}

The ranges of explosion energy $E_{\rm expl}$ and ejected $^{56}$Ni mass
$M_{\rm Ni}$ inferred from observations of light curves suggest that 
Fe core-collapse SNe fall into three categories (e.g., \cite{nomoto1}): 
(1) normal SNe with $E_{\rm expl}\approx 10^{51}$~erg
and $M_{\rm Ni}\approx 0.07\,M_\odot$, (2) hypernovae (HNe) with 
$E_{\rm expl}\sim (2$--$60)\times10^{51}$~erg
and $M_{\rm Ni}\sim 0.2$--$0.6\,M_\odot$, and (3) faint SNe with 
$E_{\rm expl}<10^{51}$~erg and $M_{\rm Ni}<0.01\,M_\odot$.
Normal SNe produce neutron stars and are associated with progenitors 
of $\leq 20\,M_\odot$. For higher masses, Fe core collapse produces
black holes and gives rise to HNe if the progenitor has sufficient
rotation to induce a jet-powered explosion or to faint SNe otherwise.
However, the detailed explosion mechanisms remain unknown in all 
three cases and different groups have employed different parametric 
approaches to model the explosion and the associated production 
and ejection of nucleosynthetic material. A commonly adopted 
parametrization is the mass cut, the material inside which is
assumed not to be ejected. If the shock is not sufficiently 
powerful to eject all the material above the mass cut, the fallback of 
some of this material results in the formation of a black hole.
In this case, the material eventually ejected contains some mixture of 
the shocked material in the inner layers and the fallback material from 
further out. This fallback and mixing requires another parametrization.

As an example, we describe the parametrization of the mass cut,
fallback, and mixing adopted in the calculations of \cite{tominaga}.
For normal SNe, the mass cut is chosen to give 
$M_{\rm Ni}=0.07\,M_\odot$. For HNe and faint SNe, in addition to
an inner mass cut $M_{\rm cut}({\rm initial})$, another parameter 
$M_{\rm mix}({\rm out})$ is introduced to set the outer boundary
of fallback. The material between the inner mass cut and this boundary
is assumed to be fully mixed. A fraction $f$ of the mixed material is
assumed to be ejected and the rest is accreted onto the black hole. 
With appropriate choices of $M_{\rm cut}({\rm initial})$,
$M_{\rm mix}({\rm out})$, and $f$, it was shown that (1) the 
overall abundance patterns of C to Zn observed in metal-poor stars 
with $-2.7<[{\rm Fe/H}]<-2.0$ can be explained by the sum of the 
contributions from normal SNe and HNe with progenitors of 
10--$50\,M_\odot$; (2) the overall patterns observed in extremely 
metal-poor stars with $-4.2<[{\rm Fe/H}]<-3.5$ can be explained by 
the ejecta from individual HNe; and (3) the overall pattern observed in 
CS~29498--043 with [Fe/H]~$\sim -3.5$ can be explained by the ejecta 
from a faint SN. However, there are also important deficiencies
of the models: the calculated abundance ratios of N, K, Sc, Ti, Mn, 
and Co relative to Fe are too low compared with observations.
As discussed in the section on the effects of rotation, primary
production of N is greatly enhanced when rotation is explicitly
included in models of metal-poor
massive stars. On the other hand, the underproduction of
K, Sc, Ti, Mn, and Co can be remedied by modifying e.g.,
the electron fraction $Y_e$ of the material undergoing
explosive nucleosynthesis.

\subsubsection{Effects of neutrinos}

The electron fraction $Y_e$ specifies the neutron-to-proton ratio 
of the material and plays a crucial role in nucleosynthesis.
The conversion between neutrons and protons can only proceed
through the weak interaction involving neutrinos. The death of
massive stars can be considered as a neutrino phenomenon.
When the Fe core of such a star collapses into a protoneutron star,
a great amount of gravitational binding energy is released in
$\nu_e$, $\bar\nu_e$, $\nu_\mu$, $\bar\nu_\mu$, $\nu_\tau$,
and $\bar\nu_\tau$ with average energies of
$\langle E_\nu\rangle\sim 10$--20~MeV. 
Typical luminosity of the initial neutrino
emission is $L_\nu\sim 10^{52}$~erg/s per species. In the case
of normal SNe, neutrino emission with $L_\nu\sim 10^{51}$~erg/s 
per species lasts for $\sim 10$~s. With such intense neutrino
fluxes, the neutron-to-proton ratio of the material close to the
protoneutron star is modified by the reactions
$\bar\nu_e+p\to n+e^+$ and $\nu_e+n\to p+e^-$. The corresponding
$Y_e$ relevant for explosive nucleosynthesis then depends on
the competition between these two reactions, which in turn depends
on the differences in $L_\nu$ and $\langle E_\nu\rangle$ between
$\bar\nu_e$ and $\nu_e$ (e.g., \cite{qian1,fuller,qwo}). 

During the first $\sim 1$~s after the onset of Fe core collapse, the
neutrino emission characteristics give rise to $Y_e>0.5$ in the
material immediately above the protoneutron star. This results in
greatly-enhanced yields of $^{45}$Sc, $^{49}$Ti, and $^{64}$Zn
due to the production of their more proton-rich progenitor nuclei
$^{45}$Cr, $^{45}$V, $^{49}$Mn, and $^{64}$Ge \cite{pruet1,frohlich1}.
It was shown that the inclusion of neutrino effects on $Y_e$ brings
the calculated abundance ratios Sc/Fe and Zn/Fe into accordance
with observations of metal-stars with $-4.1<[{\rm Fe/H}]<-0.8$
\cite{frohlich1}. As the material expands away from the protoneutron 
star, neutrinos can continue to affect nucleosynthesis. For example,
the nuclear flow in proton-rich conditions encounters bottlenecks at 
nuclei with extremely slow proton-capture 
and $\beta^+$-decay rates. In the presence of an intense $\bar\nu_e$
flux, the neutrons produced by the reaction $\bar\nu_e+p\to n+e^+$
can be captured by such nuclei to break through the bottleneck,
giving rise to the so-called $\nu p$-process \cite{frohlich2}, which
can produce many nuclei beyond $^{64}$Zn. Detailed studies of
nucleosynthesis during the expansion of proton-rich material in the
presence of intense neutrino fluxes were carried out in 
\cite{frohlich2,pruet2,wanajo2}.

\subsection{The $\alpha$-process and the $r$-process}

The reactions $\bar\nu_e+p\to n+e^+$ and $\nu_e+n\to p+e^-$ 
not only are important for determining the $Y_e$ of the material
close to the protoneutron star, but also heat this material,
enabling it to expand away from the protoneutron star as 
a neutrino-driven wind (e.g., \cite{qwo}).
In a normal SN, this wind lasts for the neutrino emission time
of $\sim 10$~s and it evolves from being proton-rich to
being neutron-rich (i.e., from $Y_e>0.5$ to $Y_e<0.5$) as
$L_{\bar\nu_e}$ and $\langle E_{\bar\nu_e}\rangle$ become
increasingly larger than $L_{\nu_e}$ and $\langle E_{\nu_e}\rangle$.
For the typical $Y_e$, entropy, and expansion timescale obtained
in the wind, the elements from Sr to Ag with mass numbers
$A\sim 90$--110 are produced through charged-particle reactions.
This is the so-called $\alpha$-process \cite{hoffman1}. If the
abundance ratio of neutrons to heavy nuclei greatly exceeds
$\sim 10$ when all charged-particle reactions cease to occur at a 
temperature of several $10^9$~K due to the Coulomb barrier, 
then the $\alpha$-process smoothly merges
with the $r$-process as the heavy nuclei produced by the former
rapidly capture neutrons at lower temperatures. This is the
neutrino-driven wind model for the $r$-process (e.g.,
\cite{woba,meyer1,takahashi,woosley,wanajo3}). However,
based on the studies by several groups \cite{qwo,witti,thompson1}, 
while it is quite plausible that the physical conditions in the wind 
are sufficient for producing the $r$-nuclei up to 
$A\sim 130$,  it is very difficult to obtain the conditions required
for producing the heavy $r$-nuclei with $A>130$ (e.g.,
\cite{meyer2,hoffman2,freiburghaus1}).

In search of the conditions for producing the heavy $r$-nuclei, 
there were attempts to modify the wind 
conditions by e.g., including the effects of a magnetic field above 
the protoneutron star \cite{thompson2} as well as proposals
of alternative sites such as
neutron star mergers (e.g., \cite{freiburghaus2}) and
the wind from the accretion disk around a black hole 
\cite{pruet3,mclaughlin}. Yet another approach is to seek 
guidance from observations. For example, the detection of
the $r$-process element ($r$-element) 
Ba with $A\sim 135$ in a number
of stars with [Fe/H]~$<-3$, especially the high Ba enrichments
in several stars with [Fe/H]~$\sim -3$, is crucial in evaluating  
neutron star mergers as the major source for the heavy 
$r$-nuclei. These events are much rarer 
(by at least a factor of $10^3$) than Fe core-collapse SNe. 
If neutron star mergers were the major source for the heavy 
$r$-nuclei, then enrichment in these nuclei would 
not occur until the ISM had already been substantially enriched in
Fe by Fe core-collapse SNe \citep{qian2,argast}. This is in contradiction 
to the observations of stars with significant to high Ba abundances
but very low Fe abundances. Therefore, it appears very unlikely
that neutron star mergers are the major source for the heavy 
$r$-nuclei.

Further guidance to the identification of the source for the heavy 
$r$-nuclei is provided by observations of a wide range
of elements from C to Zn in addition to the heavy $r$-nuclei
in a number of metal-poor stars.
Data on the metal-poor stars CS~31082--001 \cite{hill2},
HD~115444, and HD~122563 \cite{westin}  show that
their abundances of the heavy $r$-elements differ by a factor up to
$\sim 10^2$. In contrast, these stars have essentially the same
abundances of the elements between O and Ge (e.g.,
[Fe/H]~$\sim -3$).
Furthermore, when CS~22892--052 ([Fe/H]~$=-3.1$; \cite{sneden})
is compared with HD~221170 ([Fe/H]~$=-2.2$; \cite{inese})
and CS~31082--001 ([Fe/H]~$=-2.9$) with BD~$+17^\circ 3248$
([Fe/H]~$=-2.1$; \cite{cowan}), data show that
the stars in either pair have nearly the same abundances of heavy
$r$-elements but the abundances of the elements between O and Ge
differ by a factor of $\sim 8$ and 6 for the former and latter pair, respectively.
These results appear to require that the production of the heavy 
$r$-elements be decoupled from that of the elements between O and Ge
\cite{qw02,qw03,qw07}.
As Fe-core collapse SNe from progenitors of $>11\,M_\odot$ are the major
source for the latter group of elements at low metallicities, this strongly
suggests that such SNe are not the source for the heavy $r$-elements.

The elements between O and Ge are produced between the core and
the H envelope by explosive burning during a core-collapse SN or by
hydrostatic burning during its pre-SN evolution. Stars of 
$\sim 8$--$11\,M_\odot$ develop degenerate O-Ne-Mg cores, 
at least some of which eventually collapse to produce SNe
(e.g., \cite{nomoto2,nomoto3,iben}).
Models of O-Ne-Mg core-collapse SNe show that
the total amount of material ejected from between the core and the H
envelope is only $\sim 0.01$--$0.04\,M_\odot$
\cite{mayle,kitaura}, much smaller than
the $\sim 1\,M_\odot$ for Fe core-collapse SNe. Thus, O-Ne-Mg
core-collapse SNe contribute very little to the elements between O and Ge.
The decoupling between these elements and the heavy $r$-elements
can then be explained by attributing the heavy $r$-elements to 
such SNe as argued in \cite{qw02,qw03,qw07}.

The attribution of the heavy $r$-elements to O-Ne-Mg
core-collapse SNe is supported by the model proposed in
\cite{ning}. In this model, the SN shock rapidly accelerates through
the surface C-O layers of the O-Ne-Mg core due to the steep density
fall-off in these layers. This gives rise to fast
expansion of the shocked ejecta on timescales of $\sim 10^{-4}$~s.
Together with an entropy of $\sim 100$ in units of Boltzmann's
constant per nucleon and an initial electron fraction of
$Y_e\sim 0.495$ (e.g., for a composition of ${^{13}{\rm
C}}:{^{12}{\rm C}}:{^{16}{\rm O}}\sim 1:3:3$ by mass), this fast
expansion enables an $r$-process to occur in the shocked ejecta,
producing nuclei with $A>130$ through the actinides.
To further test this model requires two lines of important studies: (1)
calculating the evolution of $\sim 8$--$11\,M_\odot$ stars to
determine the pre-SN conditions of O-Ne-Mg cores, especially the
neutron excess and density structure of the surface layers; and (2)
simulating the collapse of such cores and the subsequent shock
propagation to determine the conditions of the shocked surface
layers. As these layers contain very little mass, simulations with
extremely fine mass resolutions are required to demonstrate the fast
expansion of shocked ejecta that is the key to the
production of heavy $r$-elements in the above model.

\section{Conclusions}

In summary, there have been a number of important recent
developments in theoretical and observational studies of 
nucleosynthesis, especially regarding nucleosynthetic sources 
at low metallicities. Those selected for discussion here include:
\begin{enumerate}
\item
Observations show that there is a $^6$Li plateau accompanying
the Spite plateau of $^7$Li. The corresponding $^6$Li abundances
are orders of magnitude higher than what can be produced by SBBN.
Whether this $^6$Li was produced by non-SBBN or by fusion of the
$\alpha$-particles in cosmic rays and those in the general medium 
remains to be investigated.
\item
The rather high N/O ratios observed at low metallicities indicate
contributions from rotating massive stars, in which the primary N 
production is greatly enhanced through rotationally-induced mixing.
\item
The extremely high enrichments of C and Pb observed in metal-poor 
stars of binary systems indicate that the $s$-process at low metallicities
occurs with large neutron-to-seed ratios in AGB stars and that the 
$^{13}$C pocket is the most likely neutron source. However, 
current stellar models based on the $^{13}$C pocket
substantially underproduce Ba and a number of rare-earth elements
when compared with  some observations. 
\item
While the abundance patterns of C to Zn observed in some extremely
metal-poor stars indicate contributions from individual HNe or faint SNe 
associated with progenitors of $>20\,M_\odot$, the patterns observed in 
a large number of metal-poor stars can be account for by the sum of
contributions from normal SNe and HNe. Although observations of 
metal-poor stars do not require contributions from PI-SNe associated with 
metal-free VMSs of $\approx 130$--$300\,M_\odot$, the recent report of 
possible detection of a modern PI-SN provides stimulus for further 
theoretical and observational studies of VMSs at all metallicities.
\item
The material close to the protoneutron star produced in a core-collapse SN
is subject to intense neutrino fluxes. The interaction with $\nu_e$ and
$\bar\nu_e$ changes the $Y_e$ of this material, thereby having important 
effects on the nucleosynthesis. In particular, the production of $^{45}$Sc, 
$^{49}$Ti, and $^{64}$Zn in Fe core-collapse SNe
is greatly enhanced by the proton-rich conditions of the inner ejecta
resulting from the interaction with $\nu_e$ and $\bar\nu_e$. The neutrons
produced by the reaction $\bar\nu_e+p\to n+e^+$ in proton-rich material
also give rise to a $\nu p$-process that can produce many nuclei beyond
$^{64}$Zn.
\item
The $\alpha$-process in neutron-rich neutrino-driven winds from a 
protoneutron star produced in either an Fe or an O-Ne-Mg 
core-collapse SN can easily produce 
the elements from Sr to Ag through charged-particle reactions. The
production of $r$-nuclei with $A\sim 130$ is quite likely at least in
the neutron-rich winds associated with some Fe core-collapse SNe.
However, observations require that the production of
the elements from C to Zn be decoupled from that of the heavy
$r$-nuclei with $A>130$. Therefore, the heavy $r$-nuclei must be
attributed to O-Ne-Mg core-collapse SNe
but not Fe core-collapse SNe. The rapid expansion of the shocked
surface layers of an O-Ne-Mg core may be the key to the production
of these nuclei. The SN sources for metals based on the above
discussion are summarized in Table~\ref{tab}.
\end{enumerate}

\begin{table}
\begin{tabular}{lcc}
\hline
  & \tablehead{1}{l}{b}{Fe core-collapse 
  SNe\tablenote{from progenitors of $>11\,M_\odot$}}
  & \tablehead{1}{l}{b}{O-Ne-Mg core-collapse 
  SNe\tablenote{from progenitors of $\sim 8$--$11\,M_\odot$}} \\
  \hline
C to Zn ($A<70$) & yes & no\\
Sr to Ag ($A\sim 90$--110)& yes & yes\\
$r$-process nuclei ($A\sim 130$) & yes for some? & ?\\
$r$-process nuclei ($A>130$) & no & yes\\
\hline
\end{tabular}
\caption{SN sources for metals}
\label{tab}
\end{table}

\begin{theacknowledgments}
I thank the organizers for all their efforts in making \emph{First Stars III} 
an exciting and productive conference and for inviting me. This work was
supported in part by DOE grant DE-FG02-87ER40328.
\end{theacknowledgments}

\end{document}